\documentclass[twocolumn,english,aps,pre,twocolumn,superscriptaddress,showkeys,showpacs]{revtex4-1}
\usepackage[T1]{fontenc}
\usepackage[utf8]{inputenc}
\usepackage{color}
\usepackage{babel}
\usepackage{array}
\usepackage{booktabs}
\usepackage{braket}
\usepackage{multirow}
\usepackage{amsmath}
\usepackage{amssymb}
\usepackage{graphicx}
\usepackage{float}
\PassOptionsToPackage{normalem}{ulem}
\usepackage{ulem}
\usepackage[unicode=true,pdfusetitle,
 bookmarks=true,bookmarksnumbered=false,bookmarksopen=false,
 breaklinks=false,pdfborder={0 0 1},backref=false,colorlinks=false]
 {hyperref}

\makeatletter

\hypersetup{
    colorlinks=true,
    linkcolor=red,
    citecolor=blue,
    filecolor=blue,      
    urlcolor=blue,
}
\makeatother

\begin{document}
\title{\textcolor{black}{Entropy production and efficiency enhancement in quantum Otto engines operating at negative temperatures}}

\author{Aryadine F. de Sousa}
\address{Instituto de Fí­sica, Universidade Federal de Goiás, 74.001-970, Goiânia
- GO, Brazil}
\author{Gabriella G. Damas}
\address{Instituto de Fí­sica, Universidade Federal de Goiás, 74.001-970, Goiânia
- GO, Brazil}
\author{Norton G. de Almeida}
\address{Instituto de Fí­sica, Universidade Federal de Goiás, 74.001-970, Goiânia
- GO, Brazil}
\begin{abstract}
Cyclic classical and quantum thermal machines show higher efficiency when the strokes are carried out quasi-statically. Recent theoretical and experimental work on figures of merit for thermal machines show that they have an advantage when operating in environments with negative temperatures. In an experimental proof of concept [Phys. Rev. Lett. 122, 240602 (2019)], it was shown that quantum Otto engines operating at negative temperatures can exhibit a behavior in which the faster the cycle is carried out, the higher the efficiency. In this work, we make use of the concept of entropy production and friction work to explain this counterintuitive behavior, and we show that it only occurs when reservoirs have negative temperatures.
\end{abstract}
\maketitle


\textit{Introduction}. In the 1950s, when the discussion about the existence of negative effective temperatures in spin systems began \cite{purcell1951nuclear, ramsey1956thermodynamics, abragam1958spin, abragam1957experiments,landsberg1959negative}, the manufacture of devices operating in these environments was still in its early stages of development. At that time, Zemanski's words were visionary: "Up to the present time, the only real use for systems at negative temperature has been in the rapidly expanding field of masers and lasers. Perhaps, in the future, experiments on heat mechanisms and refrigerators will be performed at negative temperatures. Then it will truly fun to be an engineer." In fact, very recent studies and experiments on the operation of devices such as heat engines in environments with negative effective temperatures show that these devices may have an advantage compared to when operating at positive temperatures. For example, thermal engines can have higher efficiency than classically expected \cite{Assis2019, xi2017quantum, mendoncca2020reservoir, nettersheim2022power, bera2024steady} and autonomous refrigerators can reach lower temperatures than would be possible if only positive temperatures were taken into account \cite{damas2023negative}.
   
Since Carnot it is well known that carrying out a cycle in finite times produces internal friction, thus increasing the  entropy  of the system,  which in turn is related to the dispersion of energy in the form of heat. Therefore, according to the common wisdom, the best way to not  waste energy by friction is to perform the strokes gently.  However, thermal machines operating in environments with negative temperatures can present unexpected and counterintuitive behaviors, such as the surprising effect that, for certain parameters, the efficiency of an Otto engine is inversely proportional to the cycle duration \cite{Assis2019}. This state of affairs raise the following questions (i) why efficiency enhances when entropy is produced? (ii) How to explain that the faster the process, the greater the efficiency? The present work aims to answer these two questions, expanding the analyzes and results in \cite{Assis2019} to clarify where the advantage generated by the physics of negative temperatures comes from.

\begin{figure}[t]
\begin{centering}
\includegraphics{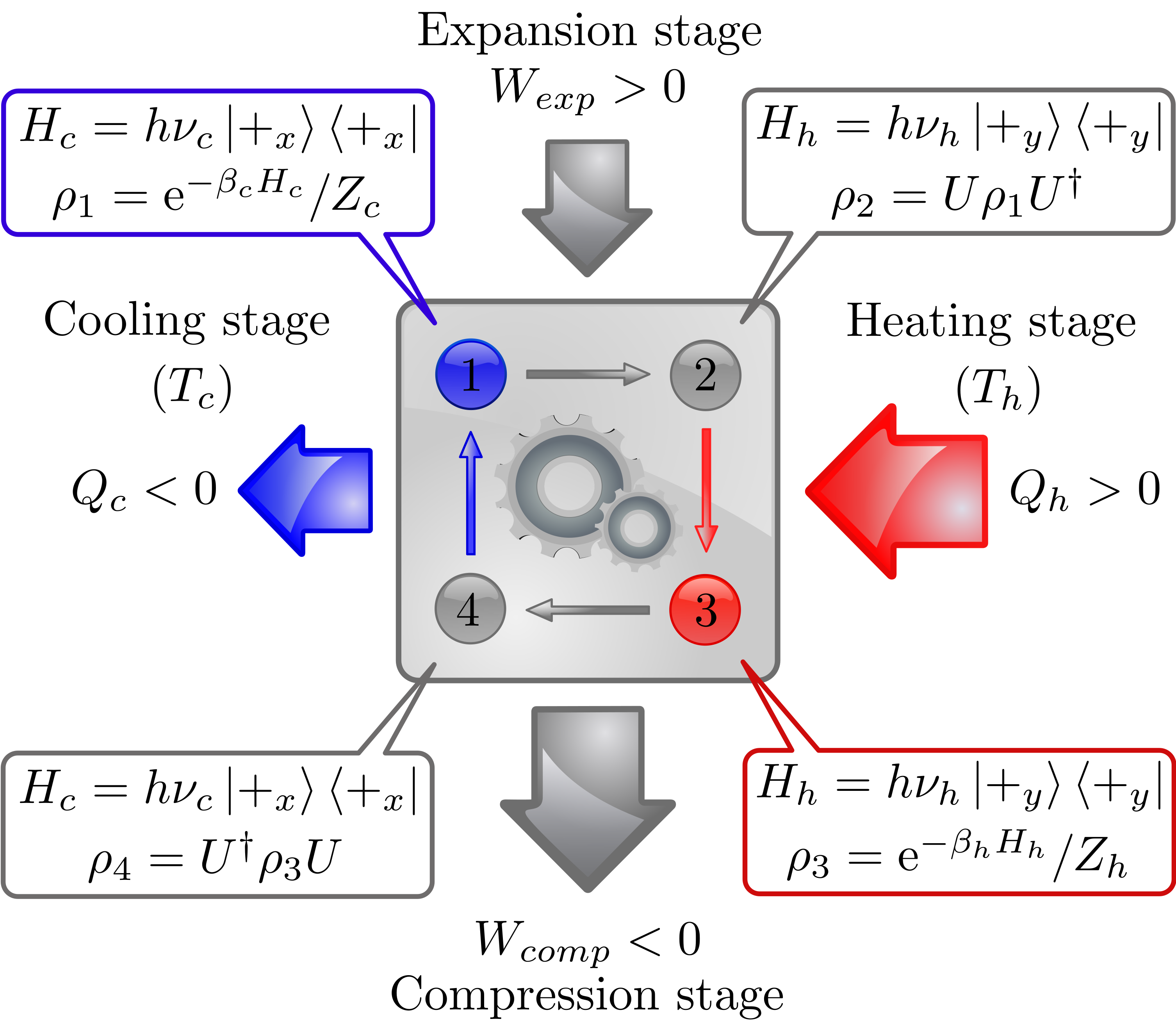}
\par\end{centering}
\caption{\label{fig:1} The four stages of a quantum Otto engine. The unitary steps occur at (1)-(2) and (3)-(4). In these steps, the working substance is isolated from the reservoirs, and its frequency is varied, resulting in positive or negative work being done, as indicated by the gray arrows vertically. The thermalization steps occur at (2)-(3) and (4)-(1). In these steps, heat exchanges occur, and the working substance is allowed to thermalize with the reservoir while its frequency is kept fixed. The heat flows occurring in these steps are shown by the red and blue arrows horizontally.}
\end{figure}

\textit{The model.} In this paper, the Quantum Otto Heat Engine (QOHE) we consider consists
of a two-level system (TLS) operating between a cold reservoir at
a positive temperature and a hot reservoir that can be either at negative  \cite{Assis2019} or 
positive \cite{peterson2019experimental} temperature.
The stages of the quantum Otto cycle that gives rise to this quantum
heat engine are as follows (see Fig. \ref{fig:1}):

\medskip{}

\noindent \emph{Expansion and compression stages.} During the expansion
and compression stages, the Hamiltonian of the TLS changes from $H_{c}=h\nu_{c}\left|+_{x}\right\rangle \left\langle +_{x}\right|$
to $H_{h}=h\nu_{h}\left|+_{y}\right\rangle \left\langle +_{y}\right|$
and $H_{h}$ to $H_{c}$, respectively, following the trajectories
dictated by $H_{exp}\left(t\right)=h\nu\left(t\right)\left[\cos\left(\pi t/2\tau\right)\left|+_{x}\right\rangle \left\langle +_{x}\right|+\sin\left(\pi t/2\tau\right)\left|+_{y}\right\rangle \left\langle +_{y}\right|\right]$
and $H_{comp}\left(t\right)=-H_{exp}\left(\tau-t\right)$, with $\nu\left(t\right)=\left(1-t/\tau\right)\nu_{c}+\left(t/\tau\right)\nu_{h}$.
Here, $h$ is the Planck constant, $\nu_{c\left(h\right)}$ $\left(\nu_{h}>\nu_{c}\right)$
is the frequency of the TLS, $\left|+_{x\left(y\right)}\right\rangle $
is the excited eigenstate of $H_{c\left(h\right)}$, and $\tau$ is
the duration of both compression and expansion stages. At the beginning
of the  expansion
and compression stages, the state of the TLS is $\rho_{1}=\text{e}^{-\beta_{c}H_{c}}/Z_{c}$
and $\rho_{3}=\text{e}^{-\beta_{h}H_{h}}/Z_{h}$, respectively, with
$\beta_{c\left(h\right)}=1/k_{B}T_{c\left(h\right)}$ and $Z_{c\left(h\right)}=\text{tr}\left(\text{e}^{-\beta_{c\left(h\right)}H_{c\left(h\right)}}\right)$,
where $k_{B}$ is the Boltzmann constant and $T_{c\left(h\right)}$
is the temperature of the cold (hot) reservoir. After evolving unitarily,
the TLS reaches the state $\rho_{2}=U\rho_{1}U^{\dagger}$ at the
end of the expansion stage and $\rho_{4}=U^{\dagger}\rho_{3}U$
at the end of the compression stage, in which $U=T_{+}\text{e}^{\left(-2\pi i/h\right)\int_{0}^{\tau}dtH_{comp}\left(t\right)}$,
where $T_{+}$ is the so-called time-ordering operator.

\medskip{}

\noindent \emph{Heating and cooling stages.} At the heating and
cooling stages, the TLS is weakly coupled to the hot and cold reservoirs,
respectively, until the equilibration process concludes. Throughout
these stages, the Hamiltonian of the TLS remains constant: $H_{h}$
during the heating stage and $H_{c}$ during the cooling stage. The
initial state of the TLS in the heating and cooling stages are, respectively,
$\rho_{2}$ and $\rho_{4}$. Upon completing the equilibration process,
the final state of the TLS is $\rho_{3}$ in the heating stage and
$\rho_{1}$ in the cooling stage.\medskip{}

According to the definitions of work and heat introduced by Alicki
in Ref. \cite{Alicki1979}, $\dot{W}=\text{tr}\bigl(\dot{H}\rho\bigr)$
and $\dot{Q}=\text{tr}\bigl(H\dot{\rho}\bigr)$, the exchange of energy
throughout the compression and expansion stages occurs in the form
of work, while it occurs in the form of heat during the heating and
cooling stages. So, by calculating the energy variation of the TLS
at each stage of the cycle ($W_{exp\left(comp\right)}=\text{tr}\left(H_{h\left(c\right)}\rho_{2\left(4\right)}\right)-\text{tr}\left(H_{c\left(h\right)}\rho_{1\left(3\right)}\right)$
and $Q_{c\left(h\right)}=\text{tr}\left[H_{c\left(h\right)}\left(\rho_{1\left(3\right)}-\rho_{4\left(2\right)}\right)\right]$),
we obtain 
\begin{equation}
W_{exp}=h\left(\nu_{h}-\nu_{c}\right)p_{c}+h\nu_{h}\xi\left(1-2p_{c}\right),\label{eq:1}
\end{equation}
\begin{equation}
W_{comp}=-h\left(\nu_{h}-\nu_{c}\right)p_{h}+h\nu_{c}\xi\left(1-2p_{h}\right),\label{eq:2}
\end{equation}
\begin{equation}
Q_{c}=-h\nu_{c}\left(p_{h}-p_{c}\right)-h\nu_{c}\xi\left(1-2p_{h}\right),
\end{equation}
and
\begin{equation}
Q_{h}=h\nu_{h}\left(p_{h}-p_{c}\right)-h\nu_{h}\xi\left(1-2p_{c}\right).
\label{eq:4}
\end{equation}
In these expressions, we are always referring to the populations of the excited states: $p_{c\left(h\right)}$ represents the population at the end of the cooling
(heating) stage, 
\begin{equation}
p_{c\left(h\right)}=\left\langle +_{x\left(y\right)}\right|\rho_{1\left(3\right)}\left|+_{x\left(y\right)}\right\rangle =\frac{1}{\text{e}^{\beta_{c\left(h\right)}h\nu_{c\left(h\right)}}+1},\label{eq:5}
\end{equation}
while $\xi$ corresponds to the transition probability between $\left|\mp_{x\left(y\right)}\right\rangle $
and $\left|\pm_{y\left(x\right)}\right\rangle $ in the compression
(expansion) stage, 
\begin{equation}
\xi=\left|\left\langle \pm_{y}\right|U\left|\mp_{x}\right\rangle \right|^{2}=\left|\left\langle \pm_{x}\right|U^{\dagger}\left|\mp_{y}\right\rangle \right|^{2},
\label{eq:6}
\end{equation}
with $\left|-_{x\left(y\right)}\right\rangle $ being the ground eigenstate
of $H_{c\left(h\right)}$. 

Upon examining Eq. (\ref{eq:5}), we can easily see that $T_{c\left(h\right)}$
is positive when $0<p_{c\left(h\right)}<1/2$ and negative when $1/2<p_{c\left(h\right)}<1$.
Therefore, a reservoir with a negative temperature inverts the population
of the TLS.

\begin{figure}[t]
\begin{centering}
\includegraphics{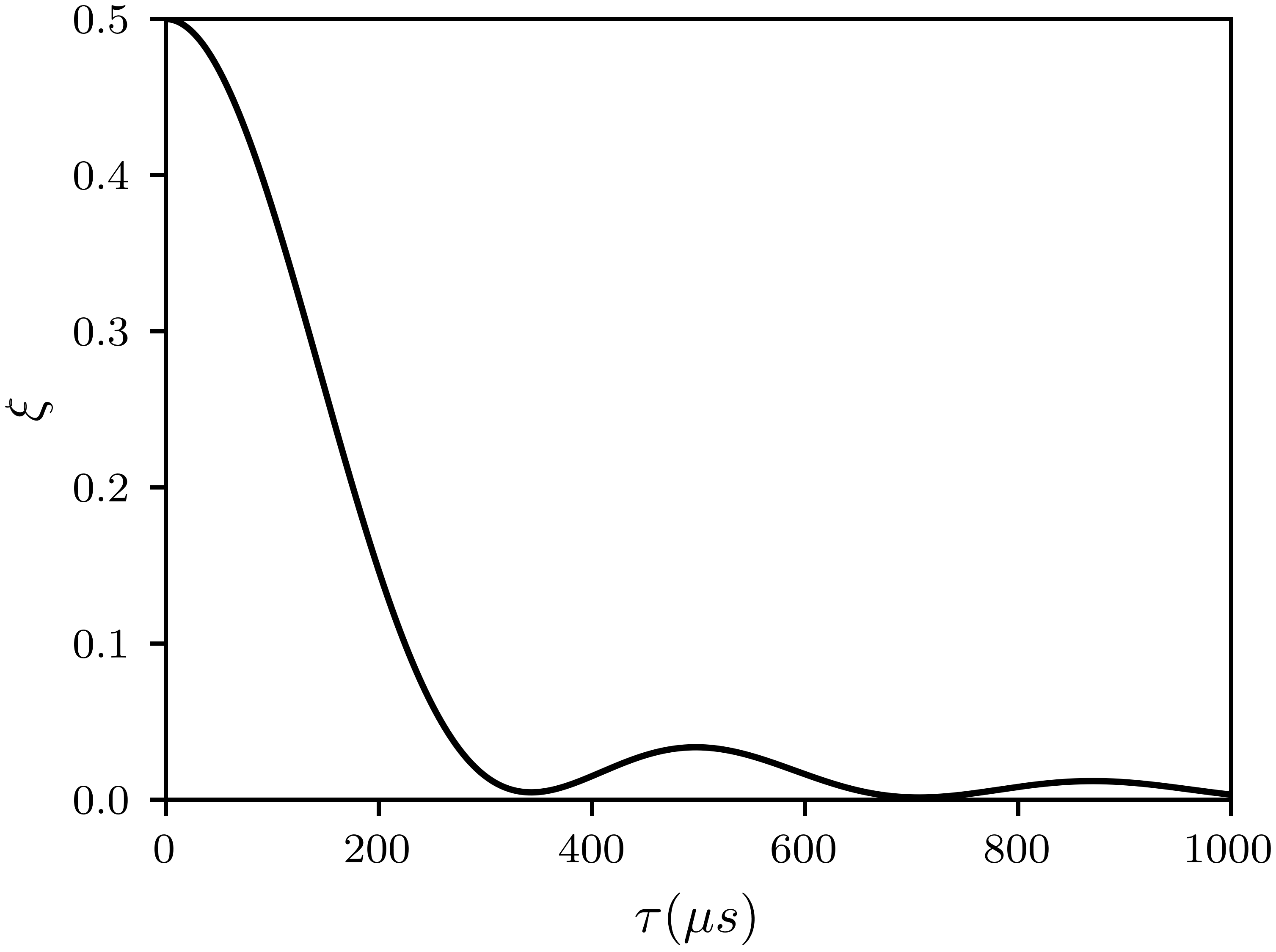}
\par\end{centering}
\caption{\label{fig:2} Transition probability $\xi$ versus expansion or compression time $\tau$. For sufficiently long times, in accordance with the adiabatic theorem, which states that no transitions occur, this parameter is null, irrespective of whether the system is in a state of population inversion. For negative temperatures, if the QOHE parameters are properly adjusted, the faster the expansion and compression processes, i.e., the smaller $\tau$, the greater the extracted work.}

\end{figure}

Due to the Hamiltonians $H_{comp}\left(t\right)$ and $H_{exp}\left(t\right)$
defined above, the transition probability $\xi$ is upper-limited
by $1/2$, reaching this limit when $U$ becomes the identity operator.
In other words, $\xi\rightarrow1/2$ as $\tau\rightarrow0$. The behavior
of $\xi$ as a function of $\tau$ is shown in Fig. \ref{fig:2},
in which we consider $\nu_{c}=2$ kHz and $\nu_{h}=3.6$ kHz. From
this figure, we can also see that $\xi\rightarrow0$ when $\tau\rightarrow\infty$,
in accordance with the quantum adiabatic theorem, regardless of whether the system is in a population-inverted state or not.

Here we will use the convention according to which the quantum Otto cycle under consideration operates as a heat engine when $W_{net}<0$, where $W_{net}$ is the network of the cycle, given
by $W_{net}=W_{comp}+W_{exp}$.  By combining Eqs. (\ref{eq:1}) and (\ref{eq:2}),
we obtain 
\begin{equation}
    \begin{split}
W_{net}=&-h\left(\nu_{h}-\nu_{c}\right)\left(p_{h}-p_{c}\right)\\
&+h\xi\left[\nu_{h}\left(1-2p_{c}\right)+\nu_{c}\left(1-2p_{h}\right)\right].
\label{eq:7}
\end{split}
\end{equation}

The work performed can be divided into two distinct parts: $W^{\mathrm{ad}}$ and $W^{\mathrm{fric}}$. The first part considers the quasi-static process, while the second part accounts for the work produced by internal friction, representing the extra work spent due to finite-time regimes. This occurs both in the presence of reservoirs at positive or negative temperatures. Therefore, the net work can be written as $W_{\text{net}} = W^{\mathrm{ad}} + W^{\mathrm{fric}}$

The work produced by internal friction can be calculated using the Kulback-Liebler divergence, which is also related to entropy production as \cite{Plastina2014}

\begin{equation}
    W^{\mathrm{fric}}_{exp(comp)}=\frac{1}{\beta_{c(h)}}D(\rho_{ 2(4)}||\rho^{qe}_{2(4)}),
    \label{eq:11}
\end{equation}
where $D$ stands for the relative entropy calculated between the states $\rho_{2(4)}$, which is the state at the end of the expansion (or compression) at finite-time, and $\rho^{qe}_{2(4)}$, which describes the resulting state when the process is conducted adiabatically. After a straightforward calculation, we get

\begin{align}
W^{\mathrm{fric}} &= h\xi\left[\nu_{h}\left(1-2p_{c}\right)+\nu_{c}\left(1-2p_{h}\right)\right]. \label{eq:10}
\end{align}

By comparing the net work given Eq. \eqref{eq:7} taking into account Eq. \eqref{eq:10} we obtain the $W^{\mathrm{ad}}$:

\begin{align}
 W^{\mathrm{ad}}=-h(\nu_h-\nu_c)(p_h-p_c). 
 \label{eq:9}
\end{align}

Notice that according to our convention, to extract work from the QOHE, we must have $W < 0$. From Eq. (\ref{eq:11}), we see that for positive temperatures the friction work is always positive, meaning that it enters Eq. (\ref{eq:7}) in a way that always reduces the useful work. On the other hand, from Eq. (\ref{eq:10}), we can see that the friction work, depending on the temperatures and operating frequencies of the engine, can contribute to either an increase or a decrease in useful work. 

For a comprehensive analysis of the efficiency of the quantum Otto cycle, since the engine condition requires that $Q_h$ be positive, and since the adiabaticity parameter $\xi$ is present in these two equations, we need to take into account the effect of this parameter on the ratio $\eta=-W_{net}/Q_h$. Let us start by studying the condition for $W^{\mathrm{fric}}$ to be negative in Eq. (\ref{eq:10}), which is

\begin{equation}
    \nu_{h}\left(1-2p_{c}\right)+\nu_{c}\left(1-2p_{h}\right)<0.
    \label{eq:12}
\end{equation}
Considering the case of negative temperatures, we observe that the populations of the excited states of the hot reservoir are contained in the interval $1/2 < p_h < 1$, while for the excited states of the cold reservoir it is $0 < p_c< 1/2$. 

Since the frequencies $\nu_h$ and $\nu_c$ are fixed, we can select an appropriate range for $p_h$ and $p_c$ to satisfy Eq.
(\ref{eq:12}). As a result, we obtain:
\begin{equation}
    \frac{1}{2}\left(1+(1-2p_{h})\frac{\nu_c}{\nu_h}\right)<p_{c}< \frac{1}{2},
    \label{eq:13}
\end{equation}
 \begin{align}
     \frac{1}{2}\left(1+(1-2p_{c})\frac{\nu_h}{\nu_c}\right)<p_{h}\leq 1.
\label{eq:14}
\end{align}

\begin{figure}[H]
	\begin{center}
		\includegraphics[scale=0.63]{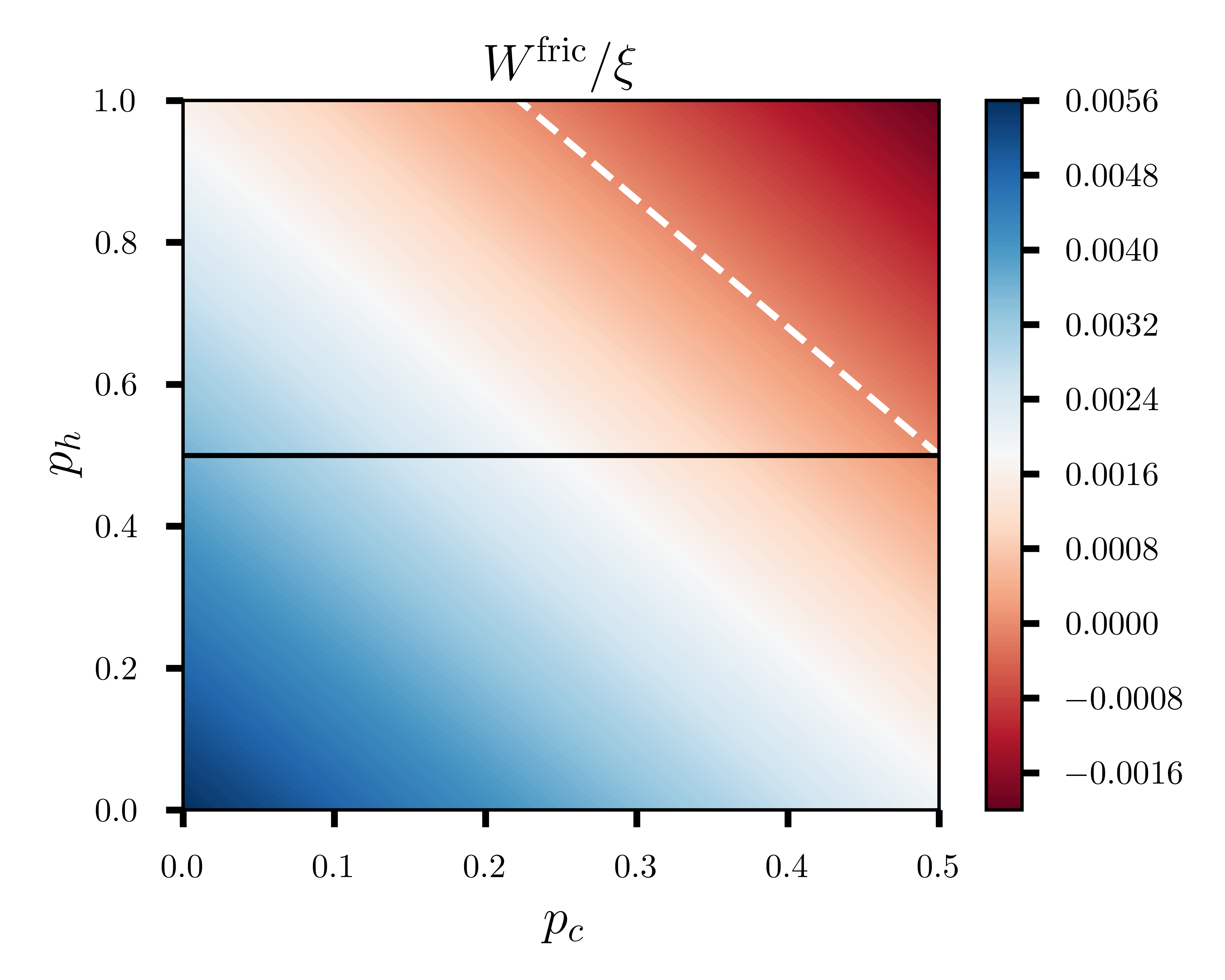}
		\caption{\label{fig:3} $W^{\mathrm{fric}}$ \emph{versus} $p_h$ and $p_c$. The populations of the hot and cold reservoirs are represented by the excited state populations $p_h$ and $p_c$, respectively. $p_h$ ($p_c$) is acquired by the system when it comes into contact with the hot (cold) reservoir and varies from 0 to 1 (0 to 0.5). The horizontal solid black line indicates the onset of population inversion ($p_h=0.5$), while the diagonal dashed white line corresponds to $W^{\mathrm{fric}}=0$. The vertical color bar next to the figure shows the scale of the friction work, which becomes more negative as $p_h$ and $p_c$ increase.}
	\end{center}
\end{figure}

To assess the impact of friction on the efficiency $\eta=-\frac{W_{\text{net}}}{Q_{h}}$, we will reframe it in relation to the populations of the excited states, based on our prior findings, Eq. (\ref{eq:7}) and Eq. (\ref{eq:4}).  Once the conditions for the cycle to operate are met—namely, $W_{\text{net}} < 0$ and $Q_h > 0$—the efficiency is:

\begin{equation}
\eta=\frac{h\left(\nu_{h}-\nu_{c}\right)\left(p_{h}-p_{c}\right)-h\xi\left[\nu_{h}\left(1-2p_{c}\right)+\nu_{c}\left(1-2p_{h}\right)\right]}{h\nu_{h}\left(p_{h}-p_{c}\right)-h\nu_{h}\xi\left(1-2p_{c}\right)},
\label{eq:16}
\end{equation}
or simply:
\begin{equation}
\eta=1-\frac{\nu_{c}}{\nu_{h}}\left[\frac{p_{h}-p_{c}+\xi\left(1-2p_{h}\right)}{p_{h}-p_{c}-\xi\left(1-2p_{c}\right)}\right].
\label{eq:17}
\end{equation}
Equation (\ref{eq:17}) enables us to examine the efficiency when the excited state populations $p_h$ and $p_c$ satisfy Eqs. (\ref{eq:13}) and (\ref{eq:14}). It is worth noting that the transition probability $\xi$ also influences the engine performance. Therefore, we can analyze the efficiency behavior for different durations of the expansion and compression stages. In cases where the process occurs adiabatically ($\xi=0$), the efficiency corresponds to $\eta_{\mathrm{ad}} = 1-\nu_c/\nu_h$, which coincides with the efficiency of the ideal quantum Otto cycle operated between two reservoirs at positive temperatures. Comparing this with Equation (\ref{eq:17}), we can observe that to have $\eta > \eta_{\mathrm{ad}}$, the following condition must be satisfied:
\begin{equation}
p_{h}>1-p_{c}.
\label{eq:18}
\end{equation}
\textit{Results}. 
Let us start by analyzing the friction work in two scenarios: with and without population inversion of the thermalized two-level system  with the reservoir at temperature $T_h$, which can be either positive or negative. In Fig. \ref{fig:3} we fixed  $\nu_c=2.0kHz$ and $\nu_h=3.6kHz$ to plot the friction work versus the population of the excited state of the hot $p_h$ and cold $p_c$ reservoirs. These are the experimental frequency values used in \cite{peterson2019experimental, Assis2019}. The population inversion occurs for $p_h > 0.5$ (above the black line), irrespective of the unitary stroke time duration. Note from Fig. \ref{fig:3} that the friction work can be both positive (below the white dashed line) and negative (above the white dashed line). When this work is null, precisely on the dashed white line, we have $W_{net}=W^{\mathrm{ad}}$, indicating that the net work corresponds to what would be obtained in the adiabatic regime, even when executing the expansion and compression steps non-adiabatically. On the other hand, when the friction work is negative, it contributes to useful work, thus improving the engine efficiency. Interestingly, the population inversion of the hot reservoir, and thus negative temperatures, allows for processes where friction is null, regardless of the time at which the unitary steps are performed as evidenced in the region below the dashed white line in Fig. \ref{fig:3} when $p_h > 0.5$.

 \begin{figure*}[t]
    \centering
    \includegraphics[width=\textwidth]{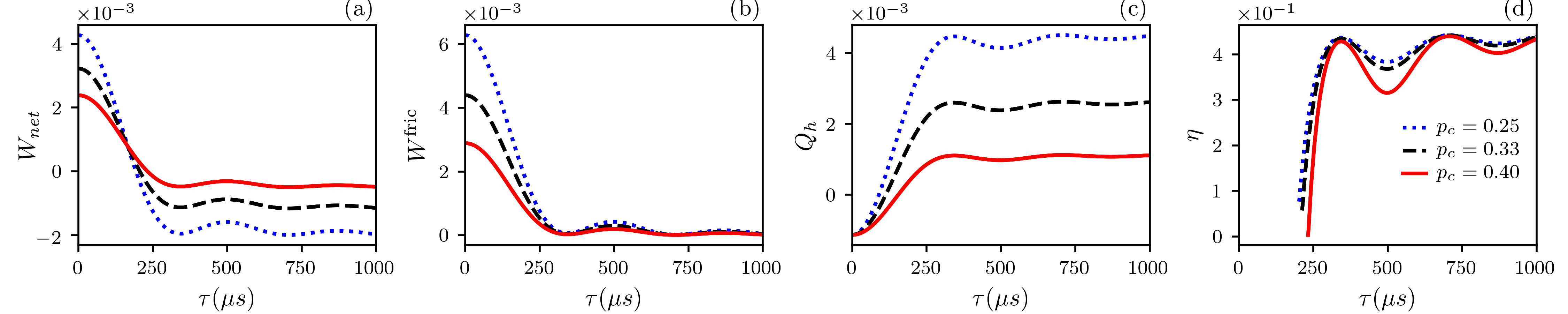}
    \caption{ A quantum Otto cycle run with positive temperatures. (a) Net work $W_{net}$ \emph{versus} the time duration $\tau$ of the unitary stroke. Note that for short times ($\tau > 200 \mu s$) the net work is positive, indicating indicating a failure to meet the engine condition. (b) Friction work $W_{fric}$  \emph{versus} $\tau$ showing that the friction work is always positive, leading to a decrease in net work. (c) Absorbed heat $Q_h$ \emph{versus} $\tau$, revealing negative values for short times, thus violating the engine condition. (d) Efficiency \emph{versus} $\tau$. Note that efficiency  initiates at $\tau > 200 \mu s$, and approaches the adiabatic limit ($\eta_{\mathrm{ad}} \approx 0.45$) as $\tau$ surpasses $700 \mu s$. }
    \label{fig:4}
\end{figure*}
 Let us now focus our attention on the net work, deepening our analysis of the Eq. (\ref{eq:7}). According to this equation, when the friction work is positive (as in the case of a cycle without population inversion), the amount of useful work extracted is smaller.
 
 \textit{Positive temperatures}. In the Fig. \ref{fig:4}\textcolor{red}{(a)} we plot the net work \textit{versus}  the time duration of the unitary stroke in a cycle run with positive temperatures. Note the absence of the engine condition for $\tau< 200 \mu s$. This absence of an engine regime occurs because, although the adiabatic work is negative in this regime, the friction work, which is larger due to entropy production at short times, prevails. In the Fig. \ref{fig:4}\textcolor{red}{(b)} we show that the friction work is always positive, thus decreasing the net work. We also studied the absorbed heat $Q_h$, as shown in Fig. \ref{fig:4}\textcolor{red}{(c)}. It can be observed that for short times, $Q_h<0$. This is justifiable because, in Eq. (\ref{eq:4}), depending on the value of $\xi$, the heat $Q_h$ can assume negative values, violating the engine regime for this time interval. The corresponding efficiency is shown in Fig. \ref{fig:4}\textcolor{red}{(d)}. Note that the efficiency starts only for $\tau>200\mu s$, which corresponds to the values of $\tau$ where the machine conditions are met. Also, observe that for values of $\tau> 700\mu s$, we have $\xi \rightarrow 0$, approaching the adiabatic limit where the efficiency tends to $\eta_{\mathrm{ad}} =1-\nu_c/\nu_h$. Given the chosen frequency values, this corresponds to $\eta_{\mathrm{ad}} \approx 0.45$.

\begin{figure*}
    \centering
    \includegraphics[width=\textwidth]{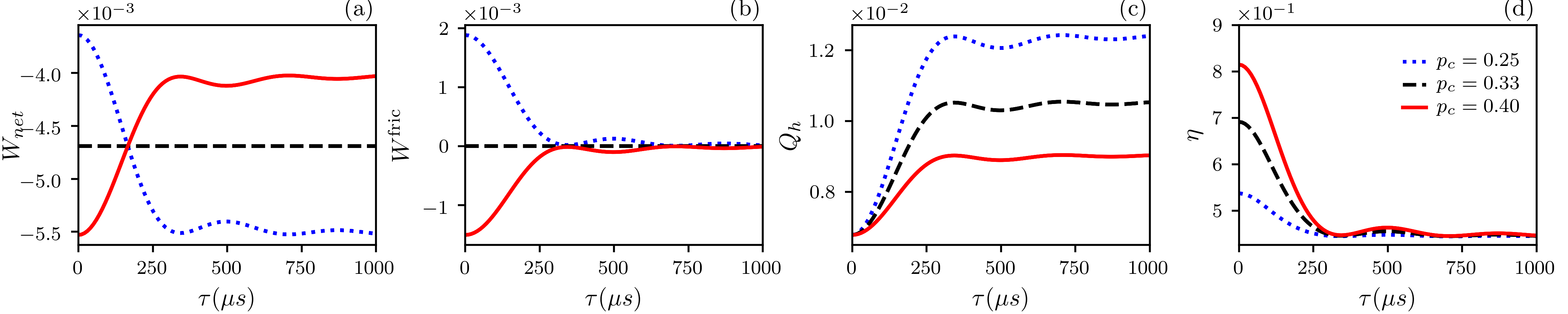}
    \caption{A quantum Otto cycle run with negative temperatures. (a) Net work $W_{\text{net}}$ versus the time duration $\tau$ of the unitary stroke. (b) Friction work $W_{\text{fric}}$ versus $\tau$, illustrating the possibility of negative or positive friction work depending on the choice of $p_c$. The dotted blue curve represents positive friction work, the solid red curve indicates negative friction work, and the dashed black curve shows zero friction work. (c) Absorbed heat $Q_h$ versus $\tau$, showing that $Q_h$ is positive for all $\tau$. (d) Efficiency $\eta$ versus $\tau$, showing an increase in efficiency for shorter times, with $\eta > \eta_{\mathrm{ad}}$ for all tested values. The solid red line represents the highest efficiency obtained, corresponding to negative friction work. Note the difference in behavior for the efficiency when compared to Fig. \ref{fig:4}(d).}
    \label{fig:5}
\end{figure*}

\textit{Negative temperatures}. In Fig. \ref{fig:5}\textcolor{red}{(a)} we show the net work versus the time duration of the unitary strokes. According to Eq.  (\ref{eq:13}), a range of populations of the excited state of the cold reservoir $p_c$ exists for which the friction work is either negative or positive. For instance, with a fixed population of $p_h=0.8$, selecting $p_c$ within the range $0.33<p_c<0.5$ results in negative friction work. Figure \ref{fig:5}\textcolor{red}{(b)} illustrates this, with the dotted blue curve representing positive friction work and the solid red curve indicating negative friction work. Additionally, the dashed black curve denotes zero friction work, independent of $\tau$. The presence of such a range where friction work is negative is significant, as it implies an algebraic addition of friction work in Eq. (\ref{eq:7}), leading to an increase in the amount of useful (negative) work extracted from the heat engine. Unlike the case of positive temperatures, the engine regime now exists for short times without restriction. When friction work is zero, the net work remains constant, corresponding solely to $W^{\mathrm{ad}}$. Notably, from the red curve, we observe that the shorter $\tau$, the more negative the net work, suggesting an increase in net work and an expected efficiency boost for short times.
Furthermore, in Fig. \ref{fig:5}\textcolor{red}{(c)}, heat $Q_h$ is depicted as a function of the unitary stroke time $\tau$. Unlike the case with a positive hot reservoir temperature, $Q_h$ is positive for all $\tau$. This, combined with the fact that $W_{\text{net}}<0$ is also negative for all $\tau$, ensures the engine condition is always met. Notably, in the regime of short times, absorbed heat progressively decreases until reaching a minimum. This occurs because the positive friction work produced in the expansion stroke, although greater for short times, leads to a subtraction of absorbed heat as per Eq. (\ref{eq:4}), resulting in increased efficiency. This efficiency increase for short times is evident in Fig. \ref{fig:5}\textcolor{red}{(d)}, where decreasing $\tau$ corresponds to increased efficiency. For all values, we observe $\eta > \eta_{\mathrm{ad}}$, indicating that $p_c$ and $p_h$ satisfy Eq. (\ref{eq:18}). The solid red line represents the highest efficiency obtained (among tested values), corresponding to the case where friction work is negative, as seen in Fig. \ref{fig:5}\textcolor{red}{(b)}.

All the analyses conducted so far help us to understand the behavior of efficiency in the regime of negative temperatures. In fact, we have seen that when one of the reservoirs has a negative temperature, the terms associated with the finite-time regime appearing in the calculation of work and heat act to increase the net work extracted from the engine at the same time to reduce the amount of heat absorbed by the engine. Since efficiency is defined as $\eta = -\frac{W}{Q_h}$, the appropriated choice for the populations $p_c$, to result in $W_\text{fric}$ always negative, allows the engine to achieve higher values for $W_{net}$ and lower values for $Q_h$, thus increasing the efficiency. Fig. \ref{fig:5}\textcolor{red}{(d)} summarizes all the previous discussions on the behavior of work $W_{net}$ and heat $Q_h$ exchanged between the working substance and the reservoir during the Otto cycle.

\textit{Conclusion}. We conducted a study of the work $W_{net}$  and the heat $Q_h$ of a quantum Otto heat engine operating in a finite time regime when one of the reservoirs presents negative temperature.  By elucidating how the negative temperature environment modifies the heat and work exchanges between the engine and the reservoir due to the production of entropy during the unitary strokes performed at finite times, we explained the recently discovered counterintuitive phenomenon that, for certain parameters of the thermal engine, the efficiency enhances when entropy is produced, and even more counterintuitive,  why the faster the process, the greater the efficiency of a heat engine.
\begin{acknowledgments}
We are grateful to Rogério J. de Assis and Celso J. Villas-Boas for insightful discussions and suggestions. We acknowledge financial support from the Brazilian agencies: Coordenação de Aperfeiçoamento de Pessoal de Nível Superior (CAPES), financial code 001. This work was performed as part of the Brazilian National Institute of Science and Technology (INCT) for Quantum Information, grant 465469/2014-0. 
\end{acknowledgments}

\bibliographystyle{apsrev4-1}
\bibliography{References}

\end{document}